\documentclass[aps,prb,amsmath,amssymb,twocolumn,showpacs,showkeys,superscriptaddress]{revtex4-1}

\usepackage{amsmath}
\usepackage[utf8]{inputenc}
\usepackage{siunitx} 
\usepackage{xcolor}
\usepackage{natbib}
\usepackage{graphicx}
\usepackage{braket}
\usepackage{comment}
\usepackage{bm}
\usepackage{epstopdf}

\begin{document}
%
%

\title{Electric field induced tuning of electronic correlation in weakly confining quantum dots}

\date{\today}

\author{Huiying Huang}
\email{huiying.emma.huang@gmail.com}
\affiliation{Institute of Semiconductor and Solid State Physics, Johannes Kepler University Linz, Altenbergerstra{\ss}e 69, A-4040 Linz, Austria}

\author{Diana Csontosov\'{a}}%
    \affiliation{Department of Condensed Matter Physics, Faculty of Science, Masaryk University, Kotl\'a\v{r}sk\'a~267/2, 61137~Brno, Czech~Republic}
    \affiliation{Czech Metrology Institute, Okru\v{z}n\'i 31, 63800~Brno, Czech~Republic}

\author{Santanu Manna}
\affiliation{Institute of Semiconductor and Solid State Physics, Johannes Kepler University Linz, Altenbergerstra{\ss}e 69, A-4040 Linz, Austria}

\author{Yongheng Huo}
\affiliation{Hefei National Laboratory for Physical Sciences at Microscale, and department of Engineering and Applied Physics, University of Science and Technology of China, Hefei, 230026, Anhui, China}

\author{Rinaldo Trotta}
\affiliation{Department of Physics, Sapienza University of Rome, Piazzale A. Moro 5, 00185 Rome, Italy}   

\author{Armando Rastelli}
\affiliation{Institute of Semiconductor and Solid State Physics, Johannes Kepler University Linz, Altenbergerstra{\ss}e 69, A-4040 Linz, Austria}    

\author{Petr Klenovsk\'{y}}%
    \email{klenovsky@physics.muni.cz}
    \affiliation{Department of Condensed Matter Physics, Faculty of Science, Masaryk University, Kotl\'a\v{r}sk\'a~267/2, 61137~Brno, Czech~Republic}
    \affiliation{Czech Metrology Institute, Okru\v{z}n\'i 31, 63800~Brno, Czech~Republic}
    

\begin{abstract}
%
%

We conduct a combined experimental and theoretical study of the quantum confined Stark effect in GaAs/AlGaAs quantum dots obtained with the local droplet etching method. In the experiment, we probe the permanent electric dipole and polarizability of neutral and positively charged excitons weakly confined in GaAs quantum dots by measuring their light emission under the influence of a variable electric field applied along the growth direction. Calculations based on the configuration-interaction method show excellent quantitative agreement with the experiment and allow us to elucidate the role of Coulomb interactions among the confined particles and even more importantly of electronic correlation effects on the Stark shifts. Moreover, we show how the electric field alters  properties such as built-in dipole, binding energy, and heavy-light hole mixing of multiparticle complexes in weakly confining systems, underlining the deficiencies of commonly used models for the quantum confined Stark effect.


%
\end{abstract}


\maketitle

\section{Introduction}
Quantum optoelectronic devices capable of deterministically generating single photons and entangled photon-pairs on demand, are considered key components for quantum photonics. Of the different available systems, semiconductor quantum dots (QDs) are one of the most promising candidates, because they combine excellent optical properties with the compatibility with semiconductor processing and the potential for scalability.~\cite{Aharonovich2016,Senellart2017,Thomas2021,Tomm2021,Orieux2017,Huber2018a,Klenovsky_IOP2010,Klenovsky2015} A prominent example is represented by GaAs/AlGaAs QDs fabricated by the local droplet etching (LDE) method~\cite{Gurioli2019,Heyn2009,Heyn2010,Huang2021,Heyn2010a,Huo2013} via molecular beam epitaxy (MBE). These QDs can show ultra-small excitonic fine-structure-splitting (FSS), with average values of $\approx4\,\mu eV$~\cite{Huo2013,Huo2014}, ultra-low multi-photon emission probabilities, with  g$^{(2)}$(0) below $10^{-4}$,~\cite{Schweickert2018}, state-of-the-art photon indistinguishabilities~\cite{Scholl2019} and near-unity entanglement fidelities of $0.978(5)$~\cite{Huber2018}. Devices based on LDE GaAs QDs have recently achieved high performance as sources of polarization-entangled photon pairs~\cite{Huber2018,Liu2019}, which led to the demonstration of entanglement swapping~\cite{Zopf2019,BassoBasset2019} and quantum key distribution~\cite{BassoBasset2021,Schimpf2021}. 
 
In addition to their excellent optical properties, semiconductor QDs 
 also provide a platform for photon-to-spin conversion~\cite{Atature2018,Borri2001}, building up bridges between photonic and spin qubits~\cite{Krapek2010}. In addition, the nuclear spins of the atoms building up a QD are emerging as long-lived quantum storage and processing units that can be interfaced to photons via coupled electron spins~\cite{Gangloff2019,Chekhovich2020}. 
To efficiently initialize and manipulate single spins confined in QDs, the QD layer is typically embedded in a diode structure, which allows the charge state to be deterministically controlled~\cite{Zhai2020}. By tuning the diode bias, not only is the charge state modified, but the magnitude of the electric field ($F_d$) along the QD growth direction is as well. In turn, $F_d$ modifies the energy and spatial distribution of the confined single particle (SP) states as well as the Coulomb and exchange interactions among the charge carriers via the so-called quantum-confined Stark effect (QCSE), leading to deep changes in the electronic and optical properties of the QDs~\cite{Patel2010,Bennett2010b, Trotta2013, Aberl2017}. Therefore, a fundamental understanding of the effects of $F_d$ in this kind of quasi-zero dimensional structures is highly desirable. 
 
LDE GaAs QDs formed by filling Al (or Ga) droplet-etched nanoholes (NHs) at high substrate temperature ($\sim600-\SI{650}{\celsius}$) present advantages over conventional strained QDs and QDs obtained by droplet epitaxy. These advantages include negligible strain, minimized intermixing of core and barrier material, a low QD density of $\approx$0.1$\,{\rm \mu m^{-2}}$, high ensemble homogeneity, and high crystal quality,~\cite{Heyn2009,Heyn2010,Heyn2014,Gurioli2019}~thus providing a particularly clean and favorable platform for both fundamental investigations and applications of QCSE.
To the best of our knowledge, only a few works have been dealing with the physics of GaAs QDs in externally applied electric fields.~\cite{Zhai2020,Singh2018,Durnev2016,Ha2015,Langer2014,Marcet2010,Ghali2012} As an example, Marcet~{\sl et~al}.~\cite{Marcet2010} and Ghali~{\sl et~al.}~\cite{Ghali2012} used vertical fields (perpendicular to the growth plane) to modify the FSS of neutral excitons confined in natural GaAs QDs (thickness or alloy fluctuations in thin quantum wells, with poorly defined density, shape and optical properties). Besides that, several simulation models based on SP assumption were also built up to explain the charge noise (emission line broadening caused by fluctuating electric field around the QDs produced by charge trapping/detrapping occurring at random places).~\cite{Heyn2020} Nevertheless, those models neither fully explain the behavior of the charge carriers in the electric field, nor take into account correlation effects~\cite{Singh2018} completely. On the contrary, we note that correlation is of particular importance in the GaAs/AlGaAs QD system because of the generally large size of the studied QDs.~\cite{Rastelli2004,Wang2009,Diana2020} For example, without including the effects of correlation, the binding energy of X$^{+}$ with respect to X$^{0}$ shall be rather small and attain negative values (anti-binding state) rather than positive ones (binding state),~\cite{Trabelsi2017} which is in contrast with the experimental observations.~\cite{Graf2014,Atkinson2012,Huber2019}
Although positive binding energies have been theoretically calculated for GaAs QDs obtained by ``hierarchical self-assembly'',~\cite{Wang2009} quantitative agreement between theory and experiment has not been demonstrated so far. In addition, detailed studies of the electric field effects on the Coulomb interactions between electrons~($e^-$) and holes~($h^+$) in GaAs QDs are still lacking. 

In this work, we conduct a combined experimental and theoretical  study of the QCSE in individual GaAs QDs. Our experiments, based on micro-photoluminescence~(\textmu-PL) spectroscopy, offer direct information on the permanent electric dipole moment (${\bf p}$) and polarizability ($\beta$)
of the neutral exciton X$^{0}$ (X$^0\equiv1e^-+1h^+$) and X$^{+}$ (X$^+\equiv1e^-+2h^+$) states in GaAs QDs, which sensitively depend on carrier interactions in those nanostructures.
In the experiment, we are able to tune the QD emission energy over a spectral range as large as 24$\,$meV thanks to the large band offsets between QD material (GaAs) and surrounding Al$_{0.4}$Ga$_{0.6}$As barriers. Such ``giant Stark effect"~\cite{Bennett2010b} allows us to observe a crossing of the X$^{+}$ emission line with that of the X$^{0}$ with increasing $F_d$, see also Appendix I.. The evolution from a binding to an anti-binding X$^{+}$ state (relative to X$^{0}$) indicates substantial electric-field-induced changes in Coulomb interactions and possibly correlation. The calculations of the aforementioned complexes are performed using the configuration-interaction (CI) method,~\cite{Shumway2001,Schliwa2009,Klenovsky2017,Klenovsky2019} see also Appendix II., with SP basis states obtained using the eight-band {\bf k}$\cdot${\bf p}~method computed with the inclusion of the full elastic strain tensor and piezoelectricity (up to second order~\cite{Bester2006,Beya2011}) by Nextnano~\cite{Birner2007} software package. Our computational approach provides consistent results with all experimental data. These calculations not only extend the investigated $F_d$'s to the range inaccessible in the experiments and explore different QD morphologies but also maps the behavior of the corresponding direct Coulomb integrals (electron-hole $J_{eh}$, hole-hole $J_{hh}$) and valence band mixing as $F_d$ is varied. Interestingly, we find that the often overlooked correlation effects among $e^-$ and $h^+$ plays a central role for describing the QCSE and that the commonly assumed quadratic dependence of the emission energy shift on $F_d$ in QDs is questionable.

\section{Quantum-Confined Stark Effect in a single G\MakeLowercase{a}A\MakeLowercase{s} QD}

\begin{figure}
	\centering
	\includegraphics[]{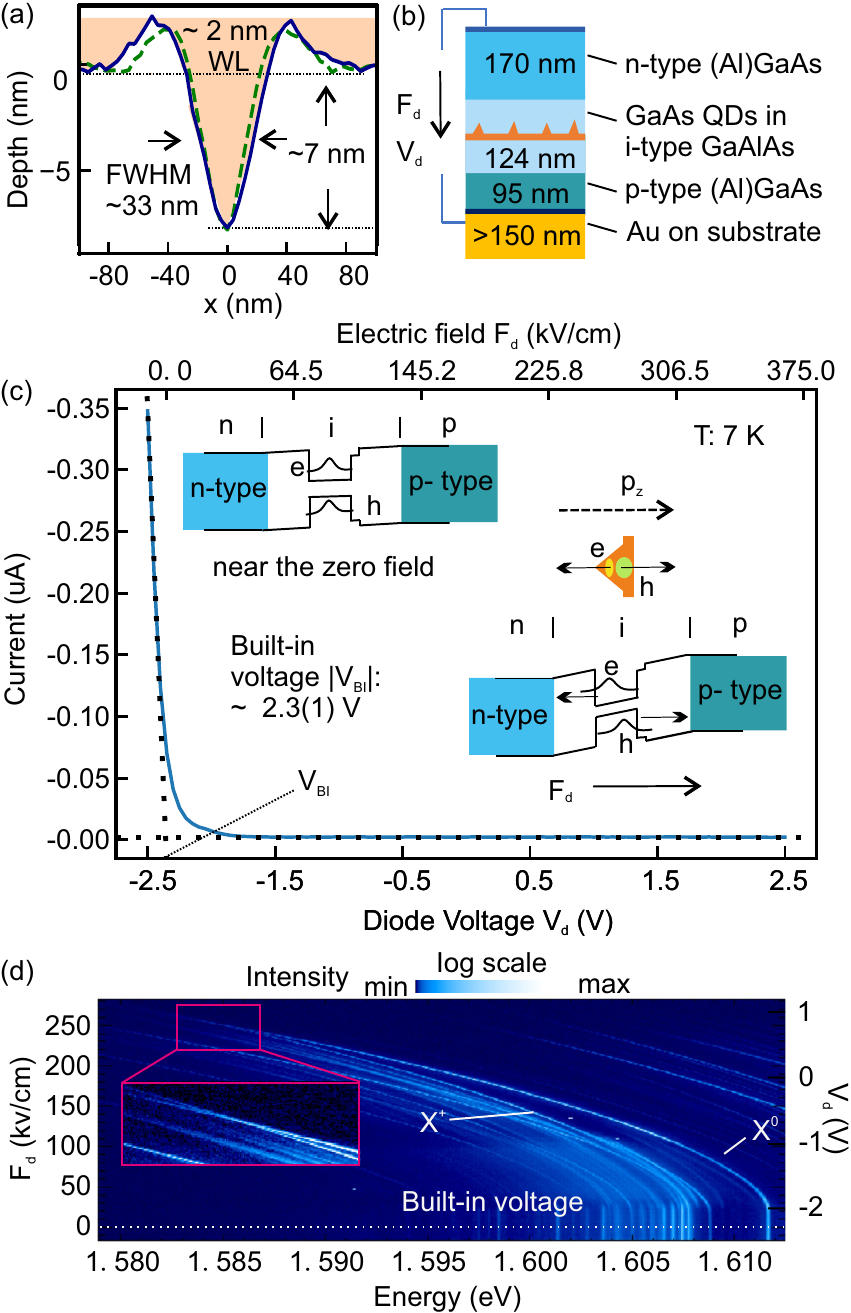}
	\caption{(a) AFM depth profile of a typical Al-droplet-etched NH. The solid and dashed lines were taken along~$[110]$~and~$[1-10]$ crystal direction of (Al)GaAs. The orange color indicates the GaAs filling and the ``wetting layer" (WL). 
		(b) Sketch of the used p-i-n Al$_{0.4}$Ga$_{0.6}$As diode with GaAs QDs in the intrinsic layer. The top and bottom of the diode membrane are protected by 10~nm of highly doped GaAs (with dimensions included within the thickness of the doped layers). The bottom Au-layer is electrically grounded. (c) The I-V characteristics of a diode at the PL measurement temperature of~$\SI{7}{\K}$. The built-in voltage ($\sim \SI{2.3}{\V}$) was estimated by the intersection of the dotted line marking the forward-bias region with the saturation current. In the inset, we show the schematic band profiles of the diode in the forward-bias and near the zero field (flat band condition). For positive $F_d$ ($F_d$ directed along the growth direction,~i.e., from the diode surface towards the gold layer), the $e^-$~($h^+$) wavefunction is pulled towards the tip~(base) of the QD. The solid and dotted arrows mark the positive direction of $F_d$ and $p_z$ respectively. (d) color-coded $\mu$-PL spectra of QD1 embedded in a p-i-n diode as a function of $F_d$ and corresponding applied voltage $V_d$. Inset: zoomed-in and intensity-enhanced part of the spectra, where we observe the crossing of X$^{0}$ and X$^{+}$.}
	\label{fig:LED}
\end{figure}
 We start by measuring the Stark shifts of X$^{0}$ and X$^{+}$ states of GaAs QDs by $\mu$-PL spectroscopy. The shape of the QD is defined by the Al-droplet-etched NH [see Fig.~\ref{fig:LED}~(a)], with a depth of~$\sim \SI{7}{\nm}$, a full width at half maximum depth of~$\sim\SI{33}{nm}$), and~$\sim 1-\SI{2}{\nano\m}$ thick ``wetting layer'' (WL) above the NHs formed by the GaAs filling.~\cite{Huo2013,Huo2014}
 
 To apply an electric field $F_d$ along the growth direction, the QDs were embedded in the intrinsic region of a p-i-n diode structure (see the details in Appendix~I)
 as sketched in Fig.~\ref{fig:LED}~(b). The direction of $F_d$ and the corresponding movement of the $e^-$ ($h^+$) wavefunction is marked in Fig.~\ref{fig:LED}~(c). $F_d$ is calculated as $F_d=(V-V_{BI})/d_i$, where $d_i$ is the thickness of the intrinsic layer ($d_i=124\,$nm) and $|V_{BI}|\simeq\SI{2.3(1)}{\V}$ is the built-in voltage of the diode [estimated from the current-voltage (I-V) trace at negative applied voltage, plotted in Fig.~\ref{fig:LED}~(c)].
 
Figure~\ref{fig:LED}~(d) shows typical $\mu$-PL spectra obtained from a QD (marked as QD1) as a function of $F_d$. Near $F_d=0$, an isolated X$^{0}$ transition is found at $1.611407(2)\,$eV, accompanied by multiexciton states at lower energies ($1.60843-1.60381\,$eV). This configuration agrees qualitatively  with other reports on GaAs QDs grown by LDE,~\cite{Huber2019,Zhai2020,Trabelsi2017} droplet epitaxy~\cite{Arashida2010} and hierarchical, self-assembly,~\cite{Rastelli2004,Wang2009} and it is different from that observed in InGaAs QDs, for which X$^{+}$ usually attains higher energy, and X$^-$ attains lower energy compared to X$^{0}$.~\cite{Regelman2001,Finley2004,Trotta2013} The X$^{0}$ state was identified by the polarization and power mapping. The other charged complexes can be calibrated by combining power mapping and temperature-dependent $\mu$-PL measurement, as shown in our previous work~\cite{Huber2019e}. Here we would like to focus only on the most intensive X$^{+}$, as it has minimal interaction (mixing) with other charged states. That X$^{+}$ was paired to the X$^{0}$ by the position check. Our sample has an ultra-low QD density ($0.3-0.4\,$QD/\textmu$m^2$), allowing single QD excitation. Energy shifts for $F_d\lesssim\SI{30}{\kV/cm}$ are not observed in our experiments because of the current injection in the diode. Investigations on the electroluminescence (EL) of this type of device have been reported previously in Ref.~\onlinecite{Huang2017}. 
The XX transition is usually not recognizable under above-band excitation (except for some values of $F_d$) due to the fact that it competes with other charged states. At large $F_d$ ($F_d\gtrsim\SI{240}{\kV/cm}$) the $\mu-$PL signal becomes faint and cannot be tracked because of the field ionization of excitons.~\cite{Finley2004} Overall, the emission energy is red-shifted by almost~24~meV upon increasing~$F_d$. We extract the energy of X$^{0}$ and X$^{+}$ by performing Gaussian fitting of their $\mu-$PL spectra for the corresponding $F_d$, and we plot those for QD1 in Fig.~\ref{fig:fitting}~(a) along with the data for another QD (marked as QD2). In both cases we observe a smaller energy shift for X$^{+}$ compared to X$^{0}$, leading to a crossing for sufficiently large values of $F_d$.

In the simulation we have modeled the NH as a cone with the basal diameter of $\SI{40}{\nm}$, a height ($h$) of $4-\SI{9.5}{\nm}$ and a wetting layer thickness of $\SI{2}{\nm}$.
Note, that later on we also provide the theory result for lens-shaped dots with the same basal diameter as reference cone-shaped dots. The lens shape, although it does not reproduce the real NH shape, has an increasing lateral space for taller QDs. In the experiments, the taller (larger) QDs will also be ``wider" than the short (smaller) one. The simulated Stark shifts of the QDs are plotted together with the experimental data from 5 dots in Fig.~\ref{fig:fitting}~(b). Calculation results are also shown for $F_d<0$, which is however not experimentally accessible with the present diode structure. It is interesting to note that the parabolic shifts are not symmetric around $F_d=0$, as already predicted in Ref.~\onlinecite{Singh2018}. Concomitantly, the maximum of the emission energy appears at $F_d>0$. Both effects are the result of the asymmetric shape of the QDs along the $F_d$~direction,~i.e., the $z$-axis combined with the different behaviors of $e^-$ and $h^+$ as their wave functions move along the $z$-axis,~thus, experiencing different lateral confinements. On the other hand, the maximum of emission energy at non-zero $F_d$ can be interpreted with the existence of a permanent electric dipole, which we will discuss in the following section. 
\section{Permanent electric dipole moments and polarizability of neutral and positively charged excitons}
\begin{figure}
	\centering
	\includegraphics[]{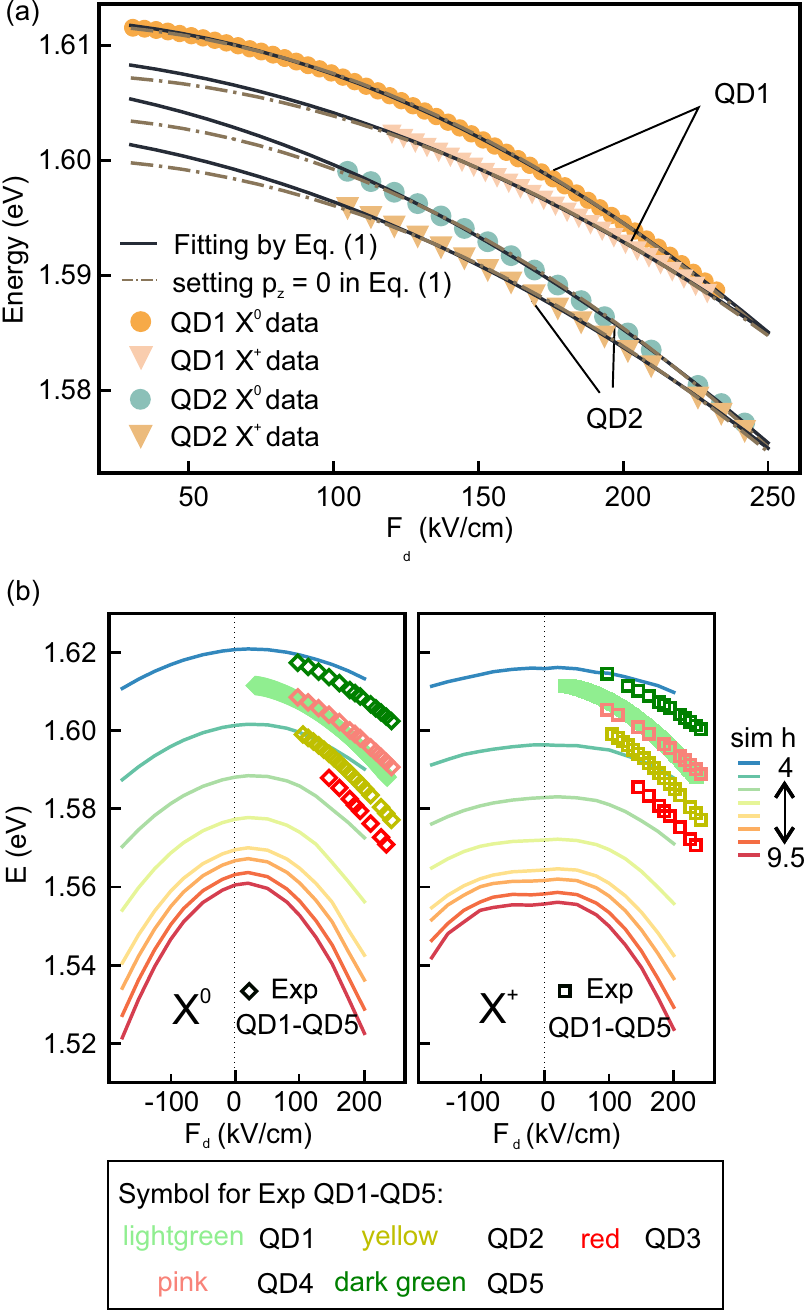}
	\caption{Measured and calculated Stark shifts of X$^{0}$ and X$^{+}$ for different QDs. (a) Experimental data for two QDs and corresponding fits using Eq.~(\ref{eq:bata}) with and without setting $p_z=0$, respectively. (b) Data for five QDs (symbols) and simulation (curves) for X$^{0}$ (left) and X$^{+}$ (right), respectively. The simulated cone-shaped dots have a fixed base diameter of $\SI{40}{\nm}$ and height varying from $4\,$nm to $9.5\,$nm. Note that the QD1 data correspond to those in Fig.~\ref{fig:LED}~(d), while data of QD2-QD5 were taken from a series of polarization resolved measurements for $F_d$ in the range of $100-\SI{250}{k\V\per\cm}$. The $\mu$-PL spectra of X$^{0}$ were fitted using Gaussian curves on one single polarization component. 
	}
	\label{fig:fitting}
\end{figure}

\begin{figure}
	\centering
	\includegraphics[]{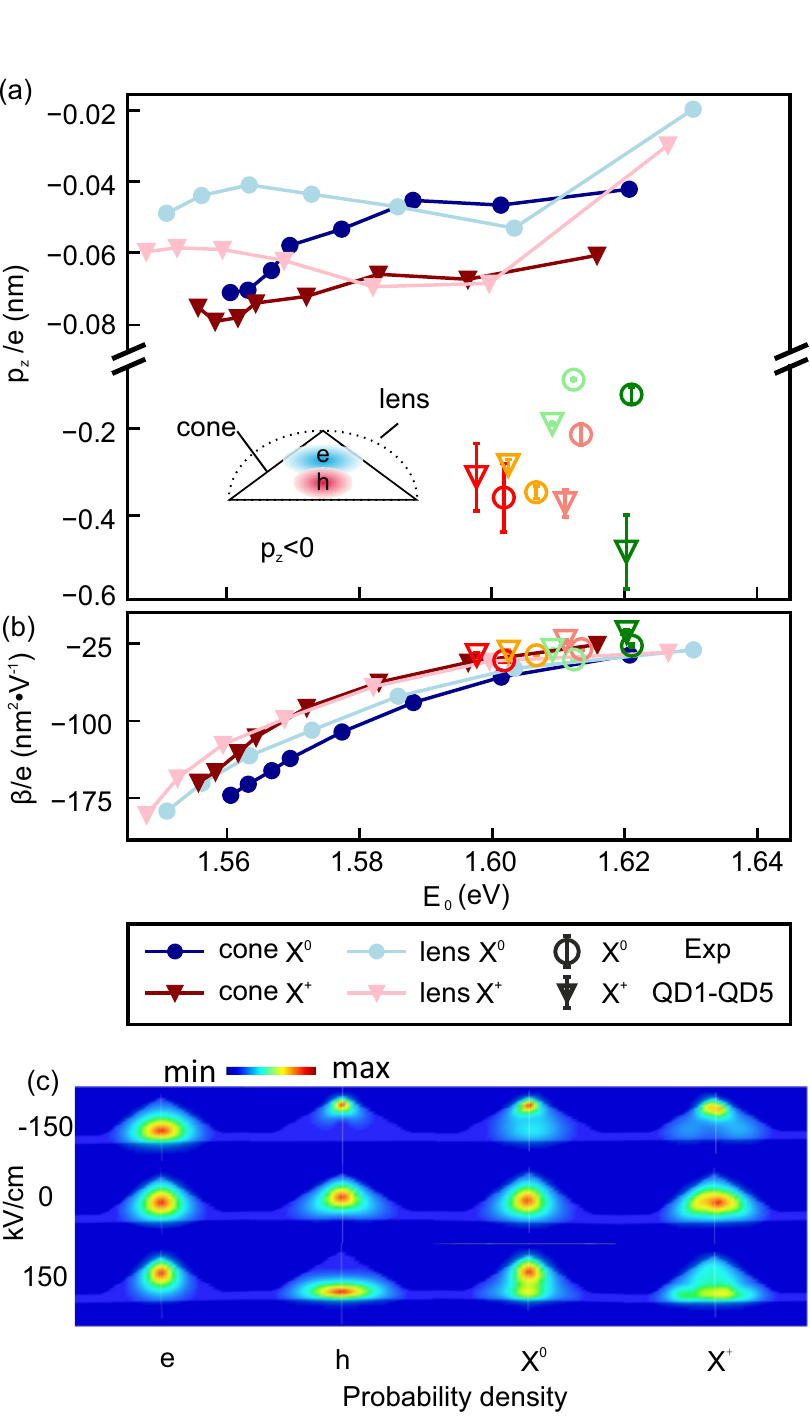}
	\caption{(a) Permanent electric dipole moments plotted as a function of the zero field energy $E_{0}$ of the corresponding complex X$^{0}$ or X$^{+}$. The parameter $p_{z}/e$ was obtained for experimental data (open symbols) by fitting the Stark shift data in Fig.~\ref{fig:fitting}~(b) by Eq.~(\ref{eq:bata}). The theoretical data of X$^{0}$ (X$^{+}$) marked by full circles (full triangles) for cone- (lens-) shaped QD are given in dark blue and dark brown (light blue and pink) and were obtained using Eq.~\eqref{eq:PzFromCI}.
	Insets: Sketch of the cone- and lens-shaped dots used in the simulation, respectively, and the corresponding position of $e^-$ and $h^+$ wavefunctions for $p_z/e<0$. Note that the height and diameter of the dot are not shown in the same scale (The dots are actually rather flat). (b) Polarizability ($\beta$) as a function of $E_{0}$. For (a) and (b): the experimental data (discrete symbols) were extracted from the Stark shift of five measured QDs in Fig.~\ref{fig:fitting}~(b) and presented in the corresponding color.
	(c) Cross-sectional view of the probability densities of $e^-$, $h^+$, X$^{0}$, and X$^{+}$ for several values of $F_d$.}
	\label{fig:beta}
\end{figure}

%
The shifts of the X$^{0}$ and X$^{+}$ energy induced by $F_d$ are commonly described by the following quadratic equation:
\begin{equation}
\label{eq:bata}
E(F_d)=E_{0}+ p_z{F_d}+\beta{F_d}^2,
\end{equation}
where $E_{0}$ is the emission energy for $F_d=0$, and $p_z$ and $\beta$ can be intuitively interpreted as the permanent electric dipole moment and polarizability of the corresponding complexes, respectively.~\cite{Jin2004,Finley2004,Aberl2017,Mar2017} The quantity $p_z/e$ can be seen as the distance between the electron and hole probability densities along the $z$-axis. The results for QD1 and QD2 fitted by Eq.~\eqref{eq:bata} for $F_d$ in the range $\SI{30}{\kV/cm}<F_d<\SI{240}{\kV/cm}$ are shown in Fig.~\ref{fig:fitting}~(a) and Table~\ref{tab:beta}. Data for X$^{+}$ at $F_d<\SI{120}{\kV/cm}$ were excluded as we could not unequivocally identify the X$^{+}$ band in that region. 
The same was done for data obtained from the other three QDs (marked as QD3-QD5 in Figures~\ref{fig:fitting}-\ref{fig:binding} and Table~\ref{tab:beta}) and the fit is performed in the $F_d$ range of $\approx100-\SI{250}{\kV/cm}$.

Figure~\ref{fig:beta}~(a) summarizes the fitted values of $p_z/e$ for X$^{0}$ and X$^{+}$ for five QDs. The negative values of $p_z/e$ for X$^{0}$ (see Table~\ref{tab:beta}) indicate that the $e^-$ wavefunction is shifted closer to the bottom of the NH (tip of the dot) compared with $h^+$ for $F_d=0$, as sketched in the bottom inset of Fig.~\ref{fig:beta}~(a). The corresponding positions of the $e^-$/$h^+$ wavefunction and the $p_z/e$ value ($p_z/e$=$-0.39$ and $\SI{-0.31}{\nm}$ for QD1 and QD2) are close to the experimental data reported in Ref.~\onlinecite{Ghali2015} and the simulated result $p_z/e$ as $\sim\SI{-0.2}{\nm}$ estimated from Fig.~4 of Ref.~\onlinecite{Heyn2020}. However, as opposed to our calculations discussed below, the computations in Ref.~\onlinecite{Heyn2020} did not consider either (i) the valence band mixing of $h^+$ or $e^-$ states and the $e^-$-$h^+$ band coupling or (ii) the correlation effects and, thus, they find negative values of $p_z/e$ only for a cone-shaped dot.
 \begin {table}
\caption {$p_z/e$ and $\beta/e$ of X$^{0}$ and X$^{+}$ from QD1 and QD2 and three more QDs identified on our sample (QD3--QD5) fitted by Eq.~(\ref{eq:bata}) }\label{tab:beta} 
\centering
\begin{tabular}{|c|c|c|c|}
	\hline 
	& $E_0$ (eV) & $p_z/e$ (nm) & $\beta/e$ (${\rm nm}^2\cdot {\rm V}^{-1}$) \\ 
	\hline 
	QD1 X$^{0}$ & 1.61234(1) & -0.082(2) & -40.36(8) \\ 
	QD1 X$^{+}$ & 1.60912(3) & -0.190(3) & -31.09(7) \\ 
	\hline 
	QD2 X$^{0}$ & 1.6067(1) & -0.34(1) & -36.15(2) \\ 
	QD2 X$^{+}$ & 1.6025(1) & -0.28(1) & -32.7(4) \\ 
	\hline 
	QD3 X$^{0}$ & 1.6018(7) & -0.36(8) & -41(2) \\ 
	QD3 X$^{+}$ & 1.5977(7) & -0.31(8) & -36(2) \\ 
	\hline 
	QD4 X$^{0}$ & 1.6135(2) & -0.21(3) & -30.1(7) \\ 
	QD4 X$^{+}$ & 1.6111(3) & -0.37(3) & -22.4(9) \\ 
	\hline 
	QD5 X$^{0}$ & 1.6211(2) & -0.12(1) & -26.8(6) \\ 
	QD5 X$^{+}$ & 1.6203(7) & -0.48(9) & -14(2) \\ 
	\hline 
\end{tabular} 
\end{table}

 We start evaluating our theoretical results for X$^{0}$ or X$^{+}$ given in Fig.~\ref{fig:fitting}~(b) by performing the same fitting procedure using Eq.~\eqref{eq:bata} as for experiment.
 However, we find that the values of $p_z/e$ obtained using that procedure depend on the range of $F_d$ where the fitting is performed. Namely, if the fitting of theoretical data by Eq.~\eqref{eq:bata} is done either for the whole range of $F_d$ values,~i.e., from $-200$ to $\SI{200}{\kV/cm}$ or just for $F_d>0$ ($F_d\in\{0-\SI{200}{\kV/cm}\}$), we find $p_z/e\in\{0-\SI{0.4}{\nano m}\}$,~i.e., {\sl positive} for most of the computed QD sizes and both considered shapes. If on the other hand, we perform the fitting for $F_d\in\{100-\SI{200}{\kV/cm}\}$,~i.e., for a similar $F_d$ range as for experiment, we find $p_z/e<0$,  in agreement with experimental data (for comparison of fits see Fig.~6 in Appendix~IV).
 Thus, the aforementioned way to 
 obtain the value of permanent electric dipole moments is unsatisfactory. It actually points to the fact that the evolution of energy of QD multi-particle complexes {\sl does not follow} equation~\eqref{eq:bata} faithfully. 
%
In order to access the intrinsic distance $p_z/e$ in GaAs QDs, we can use  
directly the SP $h^+$ and $e^-$ states, similarly to Refs.~\onlinecite{Aberl2017,Klenovsky2018}. However, this approach is reasonable only when the $e^-$-$h^+$ distance is evaluated between the SP ground states of those quasiparticles. Thus, this option is available only for X$^0$ (not X$^+$ or any complex consisting of more than two particles) and for systems that can be reasonably well described in the single-particle picture, which is not the case for GaAs/AlGaAs QDs where already X$^0$ is sizeably influenced by correlation.~\cite{Diana2020}
 Hence, instead we develop a method of obtaining ${\bf p}/e$ directly during our CI calculations~\cite{Klenovsky2017} as
\begin{equation}\label{eq:PzFromCI}
    \frac{{\bf p}^l}{e} = \sum_{m=1}^{n_{\rm SD}}{\bm \pi}_{m}|\eta^l_{m}|^2,
\end{equation}
where $\eta^l_{m}$ is an $m$-th element of the $l$-th CI matrix eigenvector $\ket{\rm M^{\mathit l}} = \left( \eta_1^l,\dots, \eta_{n_{\rm SD}}^{\mathit l} \right)^T$ corresponding to $m$-th Slater determinant (SD$_m$). Moreover, $\ket{\rm M}$ denotes the eigenstate of the CI Schr\"{o}dinger equation $H^{\rm{M}} \ket{\rm{M}} = E^{\rm{M}} \ket{\rm{M}}$, where $E^{\rm{M}}$ is the eigenenergy of that state.
Furthermore, the vector ${\bm \pi}_{m}$ relates to the following sum of all spatial integrals of $e^-$ and $h^+$ SP states corresponding to each SD$_m$ 
\begin{equation}\label{eq:PzDipole}
    {\bm \pi}_{m} = \sum_{\mathit k}\frac{\left<\Psi_{h_k}|{\bf \hat{r}}_h|\Psi_{h_k}\right>}{\left<\Psi_{h_k}|\Psi_{h_k}\right>}-\sum_{\mathit j}\frac{\left<\Psi_{e_j}|{\bf \hat{r}}_e|\Psi_{e_j}\right>}{\left<\Psi_{e_j}|\Psi_{e_j}\right>},
\end{equation}
where ${\bf \hat{r}}_h$ (${\bf \hat{r}}_e$) marks the position operator of $h^+$ ($e^-$) SP eigenstate $\left|\Psi_{h_k}\right>$ ($\left|\Psi_{e_j}\right>$), the indices $j$ and $k$ mark the SP states included in SD$_m$, and the bra-ket integrals are evaluated over the whole simulation space. Note, that in Eq.~\eqref{eq:PzFromCI} the CI eigenstates $\eta^l_{m}$ are used as ``weights" of the expectation values computed from SP states. Thus, it provides a rather general way of including the effect of correlation to the ``classical" properties related to SP states. Note that the method is partly motivated by our previous results in Ref.~\onlinecite{Diana2020}.

We show the $p_z/e$ component of Eq.~\eqref{eq:PzFromCI} in Fig.~\ref{fig:beta}~(a) for X$^0$ and X$^+$. The small computed values of $p_z/e$ -- that can be expected also from the probability density plots in Fig.~\ref{fig:beta}~(c)) (see also Appendix III.) -- are plotted together with the values (also negative) extracted by fitting the experimental data with Eq.~\eqref{eq:bata}.
The calculations indicate that the permanent electric dipole of excitons confined in GaAs QDs is very small. This is very different from the situation typically encountered in strained QDs, where the dipole is mostly determined by opposite effects, namely the alloy gradient and the strain inhomogeneities combined with piezoelectricity.~\cite{Grundmann1995,Barker2000,Fry2000,Chang1997,Findeis2001,Hsu2001,Jin2004,Sheng2001,Aberl2017}  
In view of the minuscule values of $p_z/e$ that we find in both experiment and theory it is reasonable to discard the $p_z/e$ term in fitting using Eq.~\eqref{eq:bata} in the case of our data; see also the comparison of the fitting with/without a linear term in Eq.~\eqref{eq:bata} in fig.~\ref{fig:fitting}~(a), as $|p_z/e|$ is in atomic scale

In contrast to $p_z/e$, we find for $\beta/e$ of X$^{0}$ (X$^{+}$) a more consistent agreement of fits by Eq.~\eqref{eq:bata} between theory and experiment, see Fig.~\ref{fig:beta}~(b). The results of the fits for different intervals of $F_d$ are again given in Appendix~II.
Furthermore, $\beta/e$ of X$^{0}$ (X$^{+}$) shows a clear dependence on $E_{0}$. The larger QDs, with smaller $E_{0}$, tend to have a larger magnitude of $\beta_{{\rm X}^0}$~($\beta_{{\rm X}^+}$) for X$^{0}$ (X$^{+}$), consistent with the results reported in Ref.~\onlinecite{Ghali2015}. The theoretical prediction in Ref.~\onlinecite{Barker2000,Heyn2020} also pointed out that with a fixed shape and chemical composition profile, $\beta$ is mostly sensitive to the QD height. A taller QD  provides in fact more room along the $z$-direction for the confined $e^-$-$h^+$ pairs to move away from each other when pulled apart by $F_d$, resulting in a stronger red-shift in spite of the reduced $e^-$-$h^+$ binding energy. 

We will discuss the detailed role of $e^--h^+$ Coulomb interaction and correlation in the Stark shift with the help of simulation in the following section.

\section{Trion binding energy and the role of Coulomb integrals in electric field }

\begin{figure}
	\centering
	\includegraphics[]{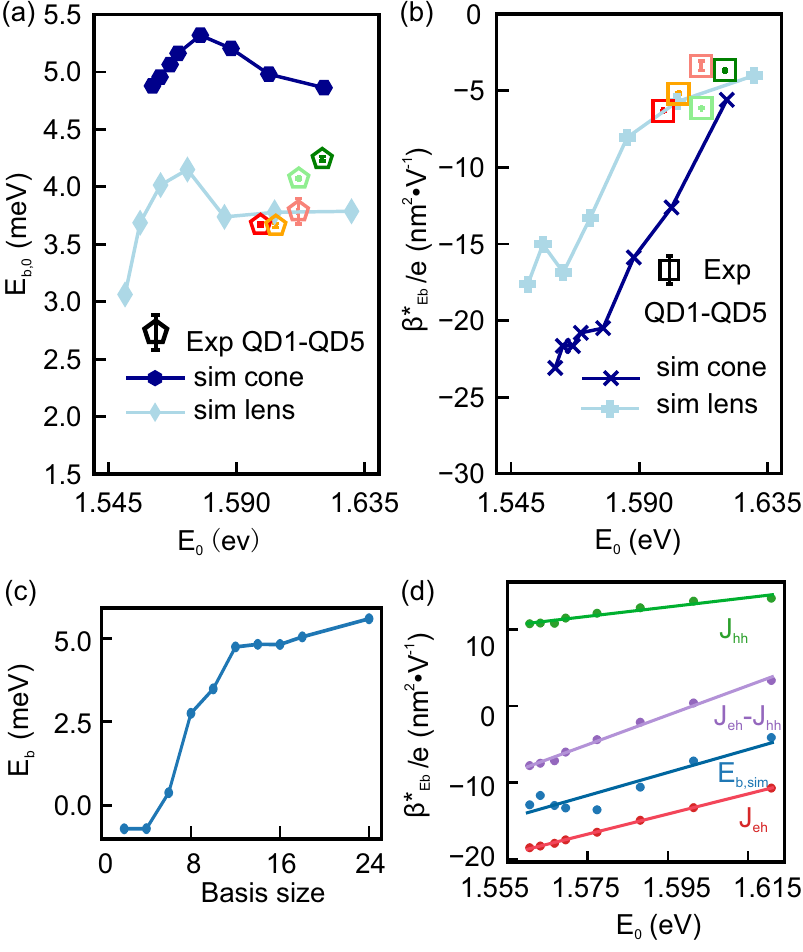}
	\caption{
	(a)~$E_{b,0}$ and (b)~$\beta^*_{E_b}/e$ as the function of $E_{0}$ fitted by Eq.~(\ref{eq:binding0}), for five QDs obtained from experiments (symbls, color marked in Fig.~\ref{fig:fitting}) and simulation (dark~blue for cone shape QDs, light~blue for lens-shape QDs). The theory values of $E_{b,0}$ in (a) were obtained directly from CI calculations, i.e., without fitting, while $\beta^*_{E_b}/e$ in (b) were obtained by fitting theory values using Eq.~\eqref{eq:binding0}. (c) Dependence of $E_b$ on the number of SP e$^-$ and SP h$^+$ states used in CI basis calculated for QD with height $h = \SI{9.5}{\nm}$. Note that we used symmetric basis,~i.e., number of SP e$^-$ states and SP h$^+$ states is equal.
	(d) Polarizabilities (full circles) of the Coulomb integrals $J_{eh}$~(red), $J_{hh}$~(green), $J_{eh}-J_{hh}$~(purple), and of $E_b$ computed by CI with 12$\times$12 SP basis~(blue). The corresponding fits by Eq.~\eqref{eq:binding0} are shown in Appendix~V.
	}
	\label{fig:binding}
\end{figure}

 To describe the evolution of the relative binding energy $E_b$ = $E$(X$^{0}$) - $E$(X$^{+}$) with $F_d$ we assume a quadratic dependence as in Eq.~\eqref{eq:bata} with an omitted linear term (see above discussion)
\begin{equation}\label{eq:binding0}
E_b(F_d)=E_{b,0}+\beta^*_{E_b}{F_d}^2,
\end{equation}
where $E_{b,0}$ marks $E_{b}$ for $F_d=0$. Thereafter, using Eq.~\eqref{eq:binding0} we fit the difference between $E$(X$^{0}$) and $E$(X$^{+}$) taken from corresponding dependencies in Fig.~\ref{fig:fitting}~(b) and we obtain the parameters $E_{b,0}$ and $\beta^*_{E_b}$, which we show alongside the calculated values in Fig.~\ref{fig:binding}~(a)~and~(b), respectively.
 From Fig.~\ref{fig:binding}~(a) we see that the calculated $E_{b,0}$ is satisfyingly close to the experimental data for both the cone- and the lens-shaped dots, in contrast to former CI calculations.~\cite{Wang2009}

Remarkably, a positive trion binding energy as large as large as 5~meV is obtained from realistic calculations. The $E_{b,0}$ values are also close to those reported in Ref.~\onlinecite{Lobl2019} ($=E_{b,0}$ linearly increasing from $\sim2.4$ to $\sim2.9$~meV for emission energies increasing from $\sim1.56$ to $\sim1.61\,$eV). We ascribe the  agreement between our theory and experiment to an almost full inclusion of the correlation effects, which will also be discussed and tested in the following.

However, we first show that the physical reason for the disagreement of Eq.~\eqref{eq:bata} with theory is due to the omission of the effect of correlation in Eq.~\eqref{eq:bata} as well.
We start by writing the energies of the final photon states after recombination of X$^{0}$ and X$^{+}$ as~\cite{Diana2020,Schliwa2008}
\begin{align}
\label{eq:coulomb_x}
E({\rm X}^0) &= \varepsilon_e-\varepsilon_h-J_{eh, {\rm X}^0}-\delta({\rm X}^0),\\
\label{eq:coulomb_x+}
E({\rm X}^+) &= \mathcal{E}_{{\rm X}^+}-|\varepsilon_h| =\varepsilon_e -\varepsilon_h-2J_{eh, {\rm X}^+}+J_{hh}-\delta({\rm X}^+),
\end{align}
where $\mathcal{E}_{{\rm X}^+}$ is the energy of X$^{+}$ before recombination, and $J_{eh, {\rm X}^0}$, $J_{eh, {\rm X}^+}$, and $J_{hh}$ are the Coulomb interactions of $e^-$-$h^+$ pairs in X$^0$ and X$^+$, and of the $h^+$-$h^+$ pair, respectively; $\varepsilon_e$ ($\varepsilon_h$) is the single particle $e^-$ ($h^+$) energy, and $\delta({\rm X}^0)$ ($\delta({\rm X}^+)$) marks the energy change due to the effect of correlation for X$^0$~(X$^+$). Consequently, the $E_b$ can be written as:
\begin{equation}\label{eq:binding}
E_b = 2J_{eh,{\rm X}^+}-J_{eh,{\rm X}^0}-J_{hh}-\delta
\end{equation}
where $\delta=\delta({\rm X}^0)-\delta({{\rm X}^+})$. Note that we have completely neglected the exchange interaction for elaborating the simplified model in Eq.~\eqref{eq:binding} since we found that to be $\approx 100$ times smaller than direct Coulomb interaction in our CI calculations (for which the exchange interaction was of course not neglected).

In Fig.~\ref{fig:binding}~(d) we plot $\beta^*_{E_b}/e$ for $J_{eh}$, $J_{eh} - J_{hh}$, $J_{hh}$, and $E_{b,sim}$ from simulation on $E_0$. Note, that $\beta^*_{E_b}/e$ values were obtained by fits using Eq.~\eqref{eq:binding0} of the theory dependencies of $J_{eh}$, $J_{eh} - J_{hh}$, $J_{hh}$, and $E_{b,sim}$ on $F_d$  computed by CI with a 12$\times$12 SP basis, for the fits see Appendix~V.
 Clearly, we find that $\beta^*_{E_b}/e$ depends on the QD size. For bigger QDs (smaller $E_0$), with steeper side facets and larger height, $|\beta^*_{E_b}/e|$ of $J_{eh}$ is more pronounced compared to that in flatter QDs. The reason is that taller QDs facilitate the $e^-$-$h^+$ separation (polarization) under the influence of vertical $F_d$. On the other hand, $|\beta^*_{E_b}|$ for $J_{hh}$ is smaller in larger QDs. The reason is that larger QDs allow the separation between $h^+$ to be larger, thus reducing the Coulomb repulsion. Since the value of $|\beta^*_{E_b}|$ for $J_{hh}$ is smaller than that of $J_{eh}$ for every QD, $\beta^*_{E_b}$ for $E_{b,sim}$ has a larger contribution of that corresponding to $J_{eh}$. 
 However, we notice that $|\beta^*_{E_b}|$ for $J_{eh} - J_{hh}$ is still smaller than that of $E_{b,sim}$ (see the corresponding curves in Fig.~\ref{fig:binding}~(c)). 
 That means, besides $J_{eh}$ and $J_{hh}$ there must be another important variable in Eq.~(\ref{eq:binding}) changing with $F_d$. Therefore, the last component in Eq.~(\ref{eq:binding}), i.e., the correlation effect $\delta$, must also vary with $F_d$ ,~i.e., $\delta=\delta(F_d)$.
 
To prove the importance of the correlation effect in our system, we calculated $E_{b}$ based on the CI model for the simulation with increasing SP basis from two $e^-$ and two $h^+$ (2$\times$2) states to twenty-four $e^-$ and twenty-four $h^+$ (24$\times$24) states. The result is plotted in Fig~\ref{fig:binding}~(c). Clearly, in the absence of correlation,~i.e.,~using 2$\times$2 and 4$\times$4 basis, X$^{+}$ is anti-binding with respect to X$^{0}$, in contradiction with the experiment. However, with increasing basis size, the effect of correlation gains importance and X$^{+}$ becomes binding with respect to X$^{0}$. The increase of $E_b$ is steep up to 12$\times$12 basis, where it almost saturates. 
Note, that the dependence was computed for the largest considered QD,~i.e.,~$h = 9.5\,\rm nm$, where the effect of correlation was expected to be the most significant.

\section{Valence band mixing of the neutral exciton and the positive trion} \label{Sec:VB_mixing}
In this section we study the effect of $F_d$ on heavy- ($\ket{\rm HH}$), light-($\ket{\rm LH}$), and spin-orbit ($\ket{\rm SO}$) hole Bloch state mixing for X$^{0}$ and X$^{+}$ ground states. The corresponding contents 
divided by the sum of those components,~i.e.,~$\varkappa(\rm HH)+\varkappa(\rm LH)+\varkappa(\rm SO)$ where $\varkappa$ marks the respective content, is shown in Fig.~\ref{fig:HH_LH}.
Note that the method of extracting the Bloch band content of CI states we show in Appendix III. (see also Ref.~\onlinecite{Diana2020}) and the conversion between \{$\ket{\rm HH}$, $\ket{\rm LH}$, $\ket{\rm SO}$\} and \{$\ket{p_x}$, $\ket{p_y}$, $\ket{p_z}$\} bases is provided in Appendix~VI.

\begin{figure}
	\centering
	\includegraphics[width=8.3cm]{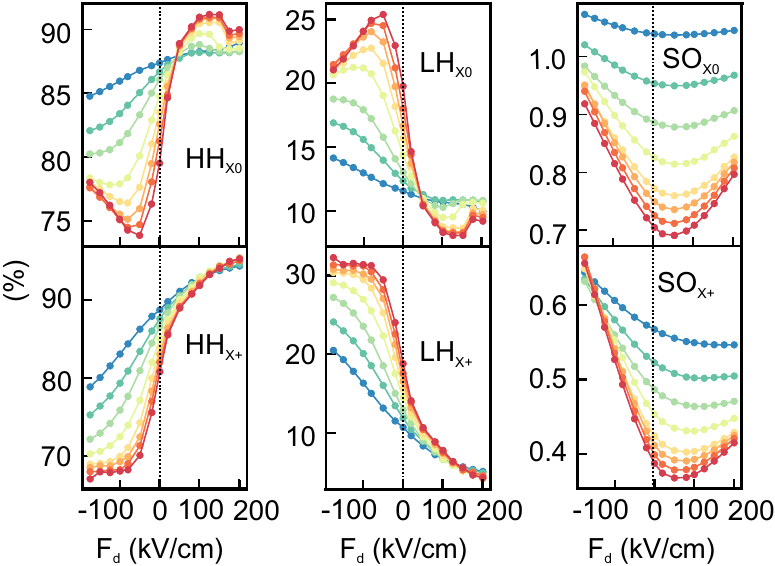}
	\caption{Contribution of $\ket{\rm HH}$, $\ket{\rm LH}$ and $\ket{\rm SO}$ states normalized to total sum of contributions of these components,~i.e., $\varkappa(HH)+\varkappa(LH)+\varkappa(SO)$, in X$^{0}$(top row) and X$^{+}$(bottom row) versus electric field $F_d$. The colors identify the heights of QDs in the same fashion as in Fig.~\ref{fig:fitting} where blue corresponds to $h = \SI{4}{nm}$ and red to $h = \SI{9.5}{\nm}$. The Bloch state contents for both X$^{0}$and X$^{+}$ were calculated using the CI model with the basis consisting of 12 SP $e^-$ and 12 SP $h^+$ state, with the effects of the direct and the exchange Coulomb interaction, and the correlation effect being included, see also Appendix III.}
\label{fig:HH_LH}
\end{figure}

We observe asymmetric dependencies around $F_d = 0$. The content of $\ket{\rm{HH}}$ increases with $F_d$ with a concomitant decrease in the contribution of $\ket{\rm{LH}}$ states.
Since the holes are pushed towards the bottom of the QD by positive $F_d$ (Fig.~\ref{fig:beta}~(c)), the $h^+$ SP state barely feels the broken translation symmetry along $z$-axis, since the lateral confinement is weaker at the bottom of the QD. Without broken symmetry the hole states tend not to mix, which causes an increase of the amount of $\ket{\rm HH}$ Bloch states. On the other hand, negative $F_d$ ($F_d$ applied along the opposite direction) pushes the holes towards the top of QD, thus, increasing the valence-band mixing (increase of the content of $\ket{\rm LH}$ and $\ket{\rm SO}$ Bloch states). According to Appendix~VI while $\ket{\rm{HH}}$ Bloch states are purely $\ket{p_x}$ and $\ket{p_y}$-like, $\ket{\rm{LH}}$ and $\ket{\rm{SO}}$ Bloch states consist also of a non-negligible amount of $\ket{p_z}$ states.
However, for the $\ket{\rm{SO}}$ states, the same amount of $\ket{p_x}$, $\ket{p_y}$, and $\ket{p_z}$ Bloch states is involved, which leads to a more symmetric trend than in the case of $\ket{\rm{HH}}$ and $\ket{\rm{LH}}$ states.

Interestingly, for negative $F_d$, the content of $\ket{\rm{HH}}$ states changes the trend after an initial decrease for $F_d$ values close to zero and starts to grow again for $F_{d} < F_{d, \rm crit}$, which is dependent on the QD height. Note, that this change is more pronounced for $\rm X^0$. Since the contents of $\ket{\rm HH}$, $\ket{\rm LH}$, and $\ket{\rm SO}$ are normalized to the total sum of all valence band components, we can directly compare X$^{0}$ and X$^{+}$. In the case of X$^{+}$ the direct and exchange Coulomb interaction between $e^-$ and $h^+$ is twice as large as that for X$^{0}$. Also the direct and exchange Coulomb interaction between two holes is included and the correlation affects the complexes in a different way; see Eqs.~(\ref{eq:coulomb_x}) and Eq.~(\ref{eq:coulomb_x+}). As one can see, the aforementioned effects influence valence-band mixing rather strongly.

Now we focus on the dot size dependence of the contents of $\ket{\rm HH}$ and $\ket{\rm LH}$ Bloch states. For $F_d< \SI{50}{\kV/cm}$ $(F_d<\SI{125}{\kV/cm})$ for X$^{0}$ (X$^{+}$), the amount of $\ket{\rm HH}$ ($\ket{\rm LH}$) Bloch states decreases (increases) with increasing height of the dot, as smaller QDs display larger energy separation between confined $\ket{\rm HH}$ and $\ket{\rm LH}$ SP states. Since the variation of valence band mixing is observed to be more pronounced in larger QDs (increased height), we observe the crossing of the HH curves for $F_d= \SI{50}{\kV/cm}$ $(F_d=\SI{125}{\kV/cm})$ in case of X$^{0}$ (X$^{+}$). Thereafter, for $F_d> \SI{50}{\kV/cm}$ $(F_d>\SI{125}{\kV/cm})$ for X$^{0}$ (X$^{+}$), the trend of the size dependence is reversed,~i.e.,~bigger QDs have a larger amount of $\ket{\rm HH}$ states than QDs with smaller height.
For such large fields the dominant part of the SP hole wavefunction leaks into the wetting layer and laterally delocalizes, leading to the a faster increase of the content of $\ket{\rm HH}$ states. We assume that for the same $F_d$ ($F_d > \SI{50}{\kV/cm}$ for X$^{0}$), all wavefunctions leak into the wetting layer with the same amount of probability density. Hence, the wavefunctions, with larger volume,~i.e.,~for bigger QDs, consist of more $\ket{p_x}$ and $\ket{p_y}$ Bloch states and so also the larger contribution of $\ket{\rm HH}$.

\section{Conclusions}
In summary, by conducting detailed $\mu$-PL spectroscopy measurements of the emission from LDE-grown GaAs/AlGaAs QDs modulated by an externally applied electric field and in conjunction with conscientious calculations of multiparticle states, we reveal the influence of the electric field on the Coulomb interaction among charge carriers in GaAs QD. The experimental data and the configuration interaction calculation clearly show the dot size dependence of the polarizability of X$^0$ and $X^+$. Thorough analysis of configuration interaction calculations sheds light on the deficiencies of the commonly used analysis of the quantum confined Stark effect by highlighting the striking effect of correlation and the direct Coulomb interaction energy between holes, which change with applied field and which are also significantly influenced by the asymmetry of the QD along the field direction, especially in large quantum dots. Moreover, we analyzed the Bloch state composition of exciton and trion complexes as a function of applied electric field, and we emphasize the influence of QD height as well. Finally, we note that our multiparticle simulation model based on the full configuration-interaction approach with large number of single-particle basis states provides excellent quantitative agreement with the experiment, and proves the non-negligible role of the correlation effect on the Stark shift for the nanosystems.

\section{Acknowledgements}
%
The authors thank A. Haliovic, U. Kainz, for technical assistance and J. Mart\'in-S\'anchez, T. Lettner for helpful discussions on the device fabrication.

This project has received funding from the Austrian Science Fund (FWF): FG 5, P 29603, P 30459, I 4380, I 4320, and I 3762, the Linz Institute of Technology (LIT) and the LIT Secure and Correct Systems Lab funded by the state of Upper Austria and the European Union's Horizon 2020 research and innovation program under Grant Agreement Nos. 899814 (Qurope), 871130 (ASCENT+).

YH Huo is supported by NSFC (Grant No. 11774326), National Key R\&D Program of China (Grant No. 2017YFA0304301) and Shanghai Municipal Science and Technology Major Project (Grant No.2019SHZDZX01).

R. Trotta is supported by the European Research council (ERC) under the European
Union’s Horizon 2020 Research and Innovation Programme (SPQRel, Grant agreement No. 679183)

D.C. and P.K. were financed by the project CUSPIDOR, which has received funding from the QuantERA ERA-NET Cofund in Quantum Technologies implemented within the European Union's Horizon 2020 Programme. In addition, this project has received national funding from the Ministry of Education, Youth and Sports of the Czech Republic and funding from European Union's Horizon 2020 (2014-2020) research and innovation framework programme under Grant agreement No. 731473.
Project 17FUN06 SIQUST has received funding from the EMPIR programme co-financed by the Participating States and from the European Union’s Horizon 2020 research and innovation programme.

\bibliography{library}

\section*{Appendix I.}
\label{appendixI}
In the experiments, the QDs were embedded in the intrinsic region of a p-i-n diode structure (thickness of~$\SI{95}{\nm}$-$\SI{124}{\nm}$-$\SI{170}{\nm}$). The thickness of the diode and the location of the QDs were chosen to obtain a simple Au-semiconductor-air planar cavity after transfer on an Au-coated substrate to enhance the out-coupling efficiency (see the details in Ref.~\onlinecite{Huang2017}). Note that, minor bi-axial strain can be introduced during processing. 

The FSS of X$^{0}$s from this sample is $\sim12-15\,\mu eV$ near zero-field and increases slightly to $\sim20\,\mu eV$ at the maximally available field due to a slight in-plane asymmetry. The linewidth of one single component of X$^{0}$ is $\sim40\,\mu eV$. The X$^{0}$ energy is chosen to be the average of the two components.  We’ve tested the consequence of choosing different polarization components. The result showed that this $\pm10\,\mu eV$ tuning has a negligible effect ($<<$ than the uncertainty) on the fitting results of $\beta$ and E$^{0}$, since $\pm10\,\mu eV$ is a quarter of X$^{0}$ linewidth and two magnitudes less than the energy difference between different dots.

In the simulation, the height of the QD is set as: 4, 5, 6, 7, 8, 8.5, 9, 9.5 for cone-shaped and 3, 4, 5, 6, 7, 8, 9 for lens-shaped in a nanometer, with a 2$\,$nm wetting layer in addition.

\section*{Appendix II.}
\label{appendixIICI}

For better readability we reproduce in Appendix II and III the description of our CI method~\cite{Klenovsky2017}, given previously also in~\cite{Diana2020}. Let us consider the excitonic complex $\ket{\rm M}$ consisting of $N_e$ electrons and $N_h$ holes. The CI method uses as a basis the Slater determinants (SDs) consisting of $n_e$ SP electron and $n_h$ SP hole states which we compute using the envelope function method based on $\mathbf{k}\cdot \mathbf{p}$ approximation using the Nextnano++ simulation suite~\cite{Birner2007}. SP states obtained from that read
\begin{equation}
    \Psi_{a_i}(\mathbf{r}) = \sum_{\nu\in\{s,x,y,z\}\otimes \{\uparrow,\downarrow\}} \chi_{a_i,\nu}(\mathbf{r})u^{\Gamma}_{\nu},
\end{equation}
where $u^{\Gamma}_{\nu}$ is the Bloch wave-function of an $s$-like conduction band or a $p$-like valence band at the center of the Brillouin zone, $\uparrow$/$\downarrow$ mark the spin, and $\chi_{a_i,\nu}$ is the envelope function, where $a_i \in \{ e_i, h_i \}$.

The trial function of the excitonic complex then reads
\begin{equation}
    \ket{\rm M} =  \sum_{\mathit m=1}^{n_{\rm SD}} \mathit \eta_m \ket{D_m^{\rm M}}, \label{eq9}
\end{equation}
where $n_{\rm SD}$ is the number of SDs $\ket{D_m^{\rm M}}$, and $\eta_m$ is the constant that is looked for using the variational method. The $m$-th SD can be found as
\begin{equation}
\ket{D_m^{\rm M}} = \frac{1}{\sqrt{N!}} \sum_{\tau \in S_N} \rm sgn \mathit(\tau) \phi_{\tau\{i_1\}}(\mathbf{r}_1) \phi_{\tau\{i_2\}}(\mathbf{r}_2) \dots \phi_{\tau\{i_N\}}(\mathbf{r}_N).
\end{equation}
Here, we sum over all permutations of $N := N_e + N_h$ elements over the symmetric group $S_N$. For the sake of notational convenience, we joined the electron and hole wave functions of which the SD is composed of, in a unique set $\{\phi_1, \dots, \phi_N\}_m := \{ \Psi_{e_j}, \dots,\Psi_{e_{j+N_e-1}}; \Psi_{h_k}, \dots,\Psi_{h_{k+N_h-1}} \}$, where $j \in \{1,\dots, n_e \}$ and $k \in \{1,\dots, n_h \}$. Accordingly, we join the positional vectors of electrons and holes $\{r_1, \dots, r_N\}:= \{ \mathbf{r}_{e_1}, \dots,\mathbf{r}_{e_{N_e}}; \mathbf{r}_{h_1}, \dots,\mathbf{r}_{h_{N_h}} \}$

Thereafter, we solve within our CI the Schr\"{o}edinger equation 
\begin{equation}
\hat{H}^{\rm{M}} \ket{\rm{M}} = E^{\rm{M}} \ket{\rm{M}} ,
\end{equation}
where $E^{\rm{M}}$ is the eigenenergy of excitonic state $\ket{\rm{M}}$, and $\hat{H}^{\rm{M}}$ is the CI Hamiltonian which reads $\hat{H}^{\rm{M}} = \hat{H}_0^{\rm{M}} + \hat{V}^{\rm{M}}$, where $\hat{H}_0^M$ represents the SP Hamiltonian and $\hat{V}^{\rm{M}}$ is the Coulomb interaction between SP states. The matrix element of $\hat{V}^{\rm{M}}$ reads~\cite{Klenovsky2017,Klenovsky2019}
\begin{equation}
\begin{split}
    &\bra{D_n^{\rm M}}\hat{V}^{\rm{M}}\ket{D_m^{\rm M}} = \frac{1}{4\pi\epsilon_0} \sum_{ijkl} \iint {\rm d}\mathbf{r} {\rm d}\mathbf{r}^{\prime} \frac{q_iq_j}{\epsilon(\mathbf{r},\mathbf{r}^{\prime})|\mathbf{r}-\mathbf{r}^{\prime}|} \\
    &\times \{ \Psi^*_i(\mathbf{r})\Psi^*_j(\mathbf{r}^{\prime})\Psi_k(\mathbf{r})\Psi_l(\mathbf{r}^{\prime}) - \Psi^*_i(\mathbf{r})\Psi^*_j(\mathbf{r}^{\prime})\Psi_l(\mathbf{r})\Psi_k(\mathbf{r}^{\prime})\}.
\end{split}
\label{eq:CoulombMatrElem}
\end{equation}
In Eq.~\eqref{eq:CoulombMatrElem} $q_i$ and $q_j$ label the elementary charge $|e|$ of either electron ($-e$), or hole ($e$), and $\epsilon(\mathbf{r},\mathbf{r}^{\prime})$ is the spatially dependent dielectric function. Note, that the Coulomb interaction is treated as a perturbation. The evaluation of the sixfold integral in Eq.~\eqref{eq:CoulombMatrElem} is performed using the Green's function method~\cite{Schliwa2009,Stier2000,Klenovsky2017,Klenovsky2019}
\begin{equation}
\begin{split}
    \nabla \left[ \epsilon(\mathbf{r}) \nabla \hat{U}_{ajl}(\mathbf{r}) \right] &= \frac{4\pi e^2}{\epsilon_0}\Psi^*_{aj}(\mathbf{r})\Psi_{al}(\mathbf{r}),\\
    V_{ij,kl} &= \int{\rm d}\mathbf{r}^{\prime}\,\hat{U}_{ajl}(\mathbf{r}^{\prime})\Psi^*_{bi}(\mathbf{r}^{\prime})\Psi_{bk}(\mathbf{r}^{\prime}),
\end{split}
\end{equation}
%
%
%
%
%
where $a,b \in \{e,h\}$ and $\nabla := \left( \frac{\partial}{\partial x}, \frac{\partial}{\partial y}, \frac{\partial}{\partial z} \right)^T$. Finally, note that $\epsilon(\mathbf{r},\mathbf{r}^{\prime})$ was set to bulk values~\cite{Klenovsky2019,Diana2020} for the CI calculations presented here.

\section*{Appendix III.}
\label{appendixIIICI}

To visualize the contents of SP states computed in multi-particle complexes calculated by CI, we need to transform the results of CI calculations to the basis of SP states instead of that of SDs.~\cite{Diana2020}

During the set-up of SDs within our CI algorithm, we create the matrix $\hat{A}$ with rank $n_{\rm SD} \times N$, where $m$-th row consists of SP states used in the corresponding SD
\begin{equation}
    A_m = \left( \Psi_{e_j}, \dots,\Psi_{e_{j+N_e-1}}; \Psi_{h_k}, \dots,\Psi_{h_{k+N_h-1}} \right).
\end{equation}

Further, resulting from diagonalization of the CI matrix, we get $n_{\rm SD}$ eigenvectors with $n_{\rm SD}$ components
\begin{equation}
    \ket{\rm M^{\mathit l}} = \left( \eta_1^l,\dots, \eta_{n_{\rm SD}}^{\mathit l} \right)^T,
\end{equation}
where the index $l$ identifies the eigenvector. We choose those values of $\eta_m^l$ that correspond to the $A_m$ consisting of a particular SP state $\Psi_{e_j}$ $\{\Psi_{h_k}\}$, we sum the squares of the absolute values  
\begin{align}
    c_{e_j} &= \sum_m \sum_{j^{\prime}} |\eta^l_{m\,(j^{\prime})}|^2 \delta_{jj^{\prime}}, \\
    c_{h_k} &= \sum_m \sum_{k^{\prime}} |\eta^l_{m\,(k^{\prime})}|^2 \delta_{kk^{\prime}},
\end{align}
and we obtain the vector
\begin{equation}
    \left( c_{e_1}^l,\dots, c_{e_{n_e}}^{l}; c_{h_1}^l,\dots, c_{h_{n_h}}^{l} \right)^T.
    \label{eqC.5}
\end{equation}
The values $c_{e_j}$ and $c_{h_k}$ are then normalized by imposing that $\sum_{j} c_{e_j}^l+\sum_{k} c_{h_k}^l = 1$. Since $|\eta_m^l|^2$ describes the weight of the corresponding SD in the CI eigenvector, we look for the weights of individual SP electron or hole states.

The procedure described thus far allows us to study also other excitonic properties, such as the influence of multi-particle effects on band mixing or visualizing the probability density of the studied excitonic complexes.

For visualizing the probability density of an eigenstate of the complex $\ket{\rm M^{\mathit l}}$ with wave-function $\Phi_{\rm{M}}^l(\mathbf{r})$ as in Fig.~\ref{fig:beta}~(c), we calculate
\begin{equation}
    |\Phi_{\rm{M}}^l(\mathbf{r})|^2 = \sum_{\mathit j} |c^l_{e_j} \Psi_{e_j} (\mathbf{r})|^2 + \sum_{\mathit k} |c^l_{h_k} \Psi_{h_k} (\mathbf{r})|^2.
\end{equation}
Finally, the probability density is finally normalized,~i.e., $\braket{\rm M^{\mathit l}|M^{\mathit l}} = 1$.

In the case of band mixing we multiply the contents of $\{\ket{\rm S}, \ket{\rm HH}, \ket{\rm LH}, \ket{\rm SO}\}$ of the particular SP state by the corresponding coefficient from Eq.~\eqref{eqC.5}. Hence, we get the matrix with rank $(n_e+n_h) \times 4$ for each $l$ and we sum separately all $\ket{\rm S}$, $\ket{\rm HH}$, $\ket{\rm LH}$ and $\ket{\rm SO}$ contents in that matrix to get the four corresponding values for each CI state. Again, we normalize the contents in the same fashion as for Eq.~\eqref{eqC.5}. The aforementioned procedure was used to obtain the results shown in Fig.~\ref{fig:HH_LH}.

\section*{Appendix IV.}
\label{appendixII}
%
%
\begin{figure}[ht!]
	\centering
	\label{fig:appendixII}
	\includegraphics[width=8.3cm]{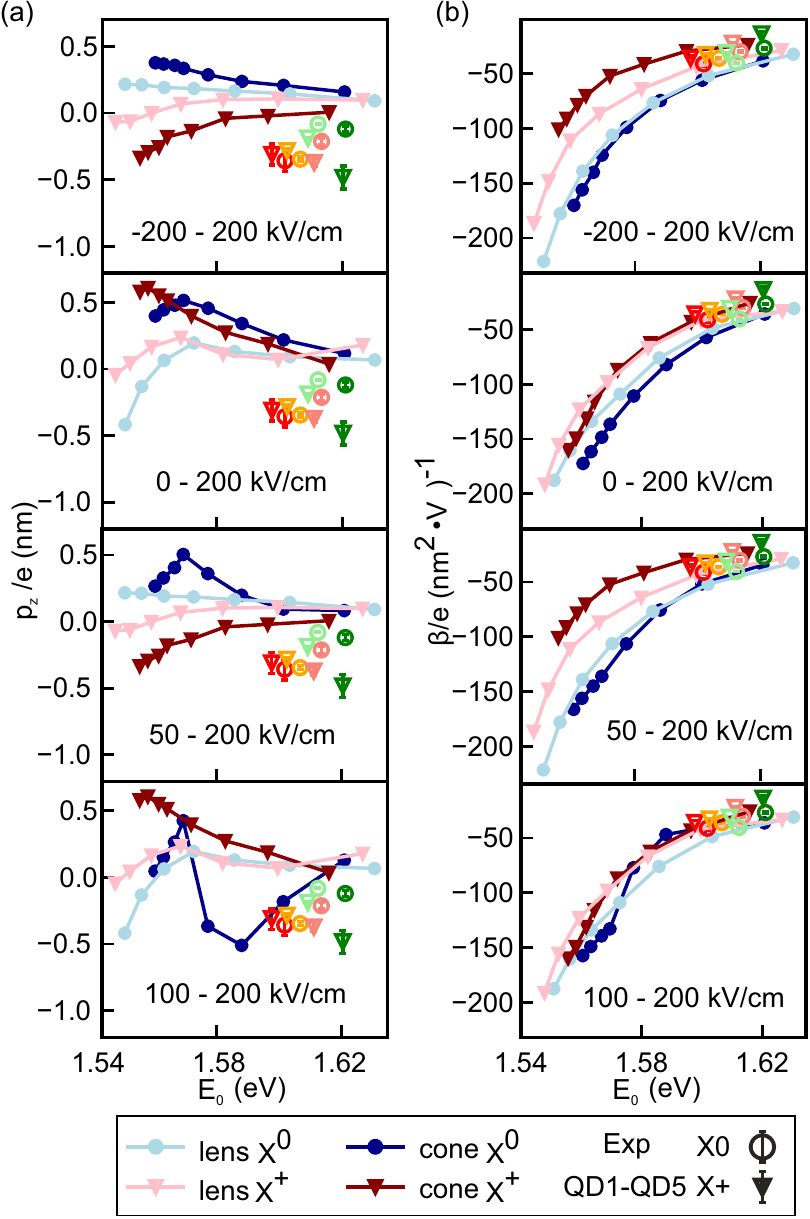}
	\caption{(a) Permanent electric dipole moments (p$_z$) and (b) Polarizability ($\beta$) plotted as a function of the zero field energy $E_{0}$ of the corresponding complex X$^{0}$ or X$^{+}$. The fits of the theoretical data (full symbols and curves) were obtained by fitting with Eq.~(1) for different ranges of $F_d$ values as indicated by the inset of each panel.}
\end{figure}

In Fig.~\ref{appendixII}, we present the permanent electric dipole moments (p$_z$) and the polarizability ($\beta$) plotted as a function of the zero field energy $E_{0}$ of the corresponding complex X$^{0}$ or X$^{+}$.
\newpage

\section*{Appendix V.}
\label{appendixIII}
\begin{figure}[ht!]
	\centering
	\includegraphics[width=3.8cm]{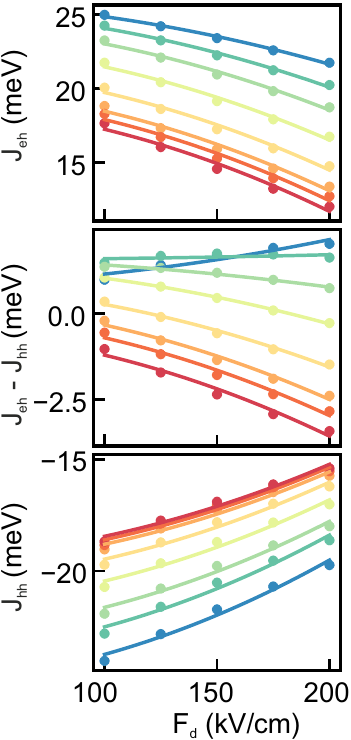}
	\caption{Dependence of $J_{eh}$, $J_{hh}$, and $J_{eh}-J_{hh}$ on $F_d$ computed by CI with 12$\times$12 SP basis. Blue data correspond to $h = \SI{4}{nm}$ and red to $h = \SI{9.5}{\nm}$.
	}
	\label{fig:binding2}
\end{figure}
In Fig.~\ref{fig:binding2}, we show the dependence of $J_{eh}$, $J_{hh}$, and $J_{eh}-J_{hh}$ on $F_d$ computed by CI with 12$\times$12 SP basis.
\newpage
\section*{Appendix VI.}
\label{appendixIV}
%
We introduce here the transformation between two $\mathbf{k}\cdot \mathbf{p}$ basis,~i.e., relation between $\{\ket{\rm S}, \ket{\rm HH}, \ket{\rm LH}, \ket{\rm SO}\} \otimes \{ \ket{\uparrow}, \ket{\downarrow}\}$ Bloch states and $\{\ket{\rm s}, \ket{\rm p_x}, \ket{\rm p_y}, \ket{\rm p_z}\} \otimes \{ \ket{\uparrow}, \ket{\downarrow}\}$ Bloch states, which has been frequently used in Section~V of the manuscript,
\begin{align}
    \ket{\rm S\uparrow} =&\,\,\bigg| \frac{1}{2},\frac{1}{2} \bigg \rangle_e = \ket{s\uparrow}, \\ 
    \ket{\rm S\downarrow} =&\,\, \bigg| \frac{1}{2},-\frac{1}{2} \bigg \rangle_e = \rm{i} \mathit{\ket{s\uparrow}}, \\ %
    \ket{\rm HH\uparrow} =&\,\, \bigg| \frac{3}{2},\frac{3}{2} \bigg \rangle = \frac{1}{\sqrt{2}} \left(\ket{p_x\uparrow} + \rm{i} \mathit{\ket{p_y\uparrow}} \right), \label{eq:HHup} \\
    \ket{\rm HH\downarrow} =&\,\, \bigg| \frac{3}{2},-\frac{3}{2} \bigg \rangle = \frac{\rm{i}}{\sqrt{2}} \left(\ket{p_x\downarrow} - \rm{i} \mathit{\ket{p_y\downarrow}} \right), \label{eq:HHdown} \\ %
    \ket{\rm LH\uparrow} =&\,\, \bigg| \frac{3}{2},\frac{1}{2} \bigg \rangle = \frac{\rm{i}}{\sqrt{6}} \left(\ket{p_x\downarrow} + \rm{i}\mathit{\ket{p_y\downarrow}} - 2\mathit{\ket{p_z\uparrow}}\right), \\ 
    \ket{\rm LH\downarrow} =&\,\, \bigg| \frac{3}{2},-\frac{1}{2} \bigg \rangle = \frac{1}{\sqrt{6}} \left(\ket{p_x\uparrow} - \rm{i}\mathit{\ket{p_y\uparrow}} + 2\mathit{\ket{p_z\downarrow}}\right), \\%
    \ket{\rm SO\uparrow} =&\,\, \bigg| \frac{1}{2},\frac{1}{2} \bigg \rangle = \frac{1}{\sqrt{3}} \left(\ket{p_x\downarrow} + \rm{i}\mathit{\ket{p_y\downarrow}} + \mathit{\ket{p_z\uparrow}}\right), \\ 
    \ket{\rm SO\downarrow} =&\,\, \bigg| \frac{1}{2},-\frac{1}{2} \bigg \rangle = \frac{\rm{i}}{\sqrt{3}} \left[ -\left(\ket{p_x\uparrow} + \rm{i}\mathit{\ket{p_y\uparrow}} \right) + \mathit{\ket{p_z\downarrow}}\right].
\end{align}
The kets $\left|J,J_z\right>$ give the total angular momentum $J$ and its projection to $z$-direction $J_z$, respectively.

\end{document}